# Reliability of Neural Networks Based on Spintronic Neurons

Eleonora Raimondo[1], Anna Giordano[1], Andrea Grimaldi[1], Vito Puliafito[2], Mario Carpentieri[3]*, Zhongming Zeng[4], Riccardo Tomasello[5], and Giovanni Finocchio[1]*

[1] *Department of Mathematical and Computer Sciences, Physical Sciences and Earth Sciences, University of Messina, I-98166, Messina, Italy*

[2] *Department of Engineering, University of Messina, I-98166, Messina, Italy*

[3] *Department of Electrical and Information Engineering, Polytechnic of Bari, via Orabona 4, 70125 Bari, Italy*

[4] *Key Laboratory of Multifunctional Nanomaterials and Smart Systems, Suzhou Institute of Nano-tech and Nano-bionics, Chinese Academy of Sciences, Ruoshui Road 398, Suzhou 215123, PR China*

[5]*Institute of Applied and Computational Mathematics, FORTH, GR-70013 Heraklion, Crete, Greece*

* *Senior Member, IEEE*

## Abstract

Spintronic technology is emerging as a direction for the hardware implementation of neurons and synapses of neuromorphic architectures. In particular, a single spintronic device can be used to implement the nonlinear activation function of neurons. Here, we present how to implement spintronic neurons with sigmoidal and rectified linear unit (ReLU)-like activation functions. We then perform a numerical experiment showing the reliability of neural networks made by spintronic neurons, all having different activation functions to emulate device-to-device variations in a possible hardware implementation of the network. Therefore, we consider a "vanilla" neural network implemented to recognize the categories of the Mixed National Institute of Standards and Technology database, and we show an average accuracy of 98.87% in the test dataset, which is very close to 98.89% as obtained for the ideal case (all neurons have the same sigmoid activation function). Similar results are obtained with neurons having a ReLU-like activation function.

Corresponding authors: Riccardo Tomasello; Giovanni Finocchio (e-mail: rtomasello@iacm.forth.gr; gfinocchio@unime.it)

## I. INTRODUCTION

The implementation of neuromorphic computing, a paradigm based on the combination of synapses and neurons, is very appealing for the potential to process large amounts of data and to perform complex operations using low-power electronic circuits. In this respect, the research has been very active both at the software level with the development of graphics processing unit based accelerators [1] and at the hardware level with the direct implementation of synaptic and neuronal functionalities with field-programmable gate arrays [2], application-specific integrated circuits [3], memristors [4], and spintronic technology. In particular, spintronic devices, such as magnetic tunnel junctions (**MTJs**), promise to reduce significantly computational costs and energy consumption for the hardware implementation of synapses and neurons [5–9]. In addition, spintronics has the advantage to offer integrability with complimentary metal-oxide-semiconductor (**CMOS**) systems, high working speeds (gigahertz or even terahertz), and intrinsic nonlinearity. The latter has been the successful property exploited in reservoir computing for recognizing spoken digits [10], as well as spoken vowels [11]. The contribution of this letter is two-fold. First, we present MTJ-based neurons having sigmoid and rectified linear unit (**ReLU**)-like activation functions (**AF**s), improving the ideas proposed in previous works [12–14]. However, when a specific software neural network architecture is transferred to the hardware level, differences in the synaptic weight and neuron AFs can occur because of the geometrical and physical device-to-device variations. Therefore, the second part of this letter studies the accuracy of a simple neural network architecture in the presence of the aforementioned variations considering neuron-to-neuron variations in the fully connected (**FC**) layer.

## II. SPINTRONIC NEURONS

*A. Sigmoidal AF*

An experimental implementation of a neuron with the sigmoid AF developed in [12] is based on a superparamagnetic MTJ. An external out-of-plane field controls the jump rate coded in an average resistance value which shows a sigmoidal behavior (see Fig. 2(d) in [12]), between two resistance states, parallel ($R_P$) or antiparallel ($R_{AP}$) depending on the relative orientations of the magnetizations of the free and polarizer layers. At zero external field, the output resistance is ($R_P$ + $R_{AP}$)/2, whereas at large-enough positive (negative) fields, it is $R_P$ ($R_{AP}$). However, a drawback of this configuration is that long times of the order of are necessary to obtain a reliable value of the output resistance. To solve this limit, here, we present a neuron with a deterministic sigmoid AF. This can be implemented using a hybrid MTJ where the free layer has a perpendicular magnetization *m* and the polarizer magnetization *p* is in-plane [see sketch in the bottom-right inset of Fig. 1(a)]. The AFs are calculated with micromagnetic computations by numerically integrating the Landau–Lifshitz–Gilbert equation [15,16]:

$$\frac{d\bm{m}}{d\tau} = -(\bm{m} \times \bm{h}_{\mathrm{eff}}) + \alpha_G \left( \bm{m} \times \frac{d\bm{m}}{d\tau} \right), \qquad (1)$$

where $\alpha_G$ is the Gilbert damping, $\bm{m} = \bm{M}/M_s$ is the normalized magnetization of the MTJ free layer, and $\tau = \gamma_0 M_s t$ is the dimensionless time, with $\gamma_0$ being the gyromagnetic ratio, and $M_s$ the saturation magnetization. $\bm{h}_{\mathrm{eff}}$ is the normalized effective magnetic field, which includes the exchange, perpendicular anisotropy, magnetostatic and external fields.

We analyze an elliptical 20 nm×18 nm×1 nm MTJ with a discretization cell of 1 nm×1nm×1 nm. The nominal physical parameters are $M_S$ = 1100 kA/m, exchange constant $A$ = 20 pJ/m, and $\alpha_G$ = 0.02.

At zero external field, the free layer magnetization is out-of-plane (perpendicular anisotropy constant $K_u$ = 0.6215 MJ/m$^3$). Therefore the $x$-component is zero, and so it is the output resistance $R(\theta) = R_P + (R_{AP} - R_P)[1 - \cos(\theta)]/2$, where $\theta$ is the angle between free layer and polarizer magnetization vectors. An external field $H_x$ applied along the $x$-axis controls the equilibrium direction of the magnetization, which will eventually align with $H_x$. The simulation results are shown in the main panel of Fig. 1(a) (red circles) and are well-fitted by an ideal sigmoid $\sigma(x) = \frac{1}{1 + e^{-\alpha x}}$ with $\alpha = 1$, thus confirming that this MTJ configuration can be used as a deterministic neuron with a sigmoid AF. $H_x$ can be generated by a dc current flowing in a nanowire on top of the MTJ, as sketched in the bottom-right area of Fig. 1(a). Specifically, in our model, a positive (negative) field is generated by a negative (positive) current $I_{DC}$, therefore promoting the parallel (antiparallel) state.

To study the effect of device-to-device variations in implementing the FC layer, we have performed simulations considering a maximum variation of the anisotropy constant ranging from $K_u$ = 0.6200 to 0.6225 MJ/m$^3$. These values are the lowest ones that ensure a perpendicular free layer at zero temperature, as well as differing of about 5%, which is an expected difference when depositing MTJs with the same nominal thickness. The previous anisotropy values still yield a sigmoid AF with coefficient $\alpha = 0.91$ and $\alpha = 1.30$, respectively, as shown in the top-left inset of Fig. 1(a). The use of a larger $K_u$ while maintaining the other parameters fixed ensures a stable perpendicular easy-axis also in the presence of thermal fluctuations at room temperature. In this case, we still find a sigmoidal AF (not shown) which is linked to the time-averaged value of the magnetization $x$-component. This averaged value can be considered reliable only if the stochastic signal due to thermal fluctuations is sufficiently long. However, by calculating the moving mean of the magnetization x-component, we observed that the time-averaged magnetization tends to a constant value already after 5 ns (not shown), which is still a very small time.

We wish to highlight that it is possible to control the anisotropy by finely tuning it with a voltage applied across the MTJ [17–20]. This is the reason we are considering small variations. This aspect must be considered important; otherwise the neurons can work at very different input fields.

An alternative implementation of a deterministic neuron with a sigmoid AF can also be realized by taking advantage of the spin-transfer torque [21,22] due to a spin-polarized current [see Fig. 1(b)]:

$$\tau_{\text{OOP}} = \frac{g\mu_B J_{\text{DC}}}{\gamma_0 e M_S^2 L_{\text{FM}}} \frac{2\eta}{\left[1+\eta^2 (\mathbf{m}\cdot\mathbf{p})\right]} \left[\mathbf{m}\times(\mathbf{m}\times\mathbf{p})\right] \quad (2)$$

with $g$ being the Landé factor, $\mu_B$ the Bohr magneton, $J_{\text{DC}}$ the current density flowing through the MTJ, $e$ the electron charge, $L_{\text{FM}}$ the thickness of the free layer, and $\eta = 0.66$ the spin polarization.

This solution does not require an external field at the cost of a larger area occupancy. Indeed, it is necessary to use two hybrid MTJs in parallel with two diodes that select which MTJ to bias depending on the current sign [see inset in Fig. 1(b)]. This dual-branch scheme is needed because for one sign of the current, either a self-oscillation state or switching dynamics is excited depending on the current amplitude [23–26].

*B. ReLU-Like AF*

Spintronic diodes have been demonstrated to be very promising for microwave detection [27–30] overcoming the thermodynamics limit of Schottky diodes. In addition, those devices can be used for neuromorphic applications. For example, in [13], it has been shown how to implement a neuron with a ReLU-like AF. The key aspect is to consider the ascending branch of the rectified signal obtained as a function of $I_{\text{DC}}$ (see Fig. 3(a) in [13]).

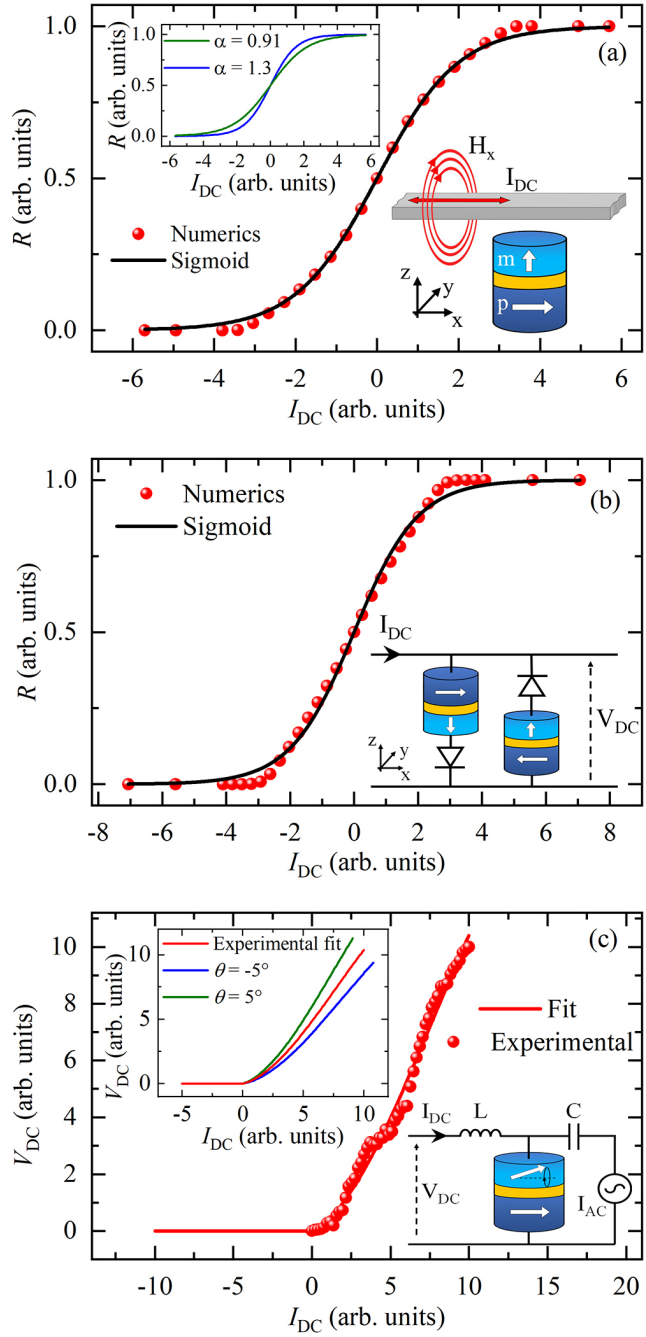

Fig. 1. (a) Example of a magnetoresistance curve of the MTJ device as a function of $I_{DC}$ as computed from micromagnetic simulations for $K_u$ = 0.6215 MJ/m$^3$ (red dots) fitted by the ideal sigmoid function (black line). (Top-left inset) Magnetoresistance curves of the MTJ device for $K_u$ = 0.6225 MJ/m$^3$ and 0.6200 MJ/m$^3$, which can be described by a sigmoid with a slope coefficient α = 0.91 and 1.3, respectively. (Bottom-right inset) A sketch of the deterministic neuron sigmoid AF composed of an MTJ and a current line. The $I_{DC}$ current generates the in-plane magnetic field, which aligns the free layer magnetization **m** parallel or antiparallel to the polarizer magnetization **p**. (b) Magnetoresistance curve of the neuron concept involving two MTJs and two diodes (inset). The red dots represent the micromagnetic simulations' results fitted by an ideal sigmoid function (solid black line). (c) Experimental rectified voltage $V_{DC}$ as a function of dc current. The red dots represent the experimental data from Ref. [31] fitted by a third-order polynomial function (red line). We cannot increase the polynomial order beyond four because the fitting curve becomes nonmonotonic. (Top-left inset) Fitted experimental data with different spreading angles, as indicated in the legend. (Bottom-right inset) Sketch of the MTJ spintronic diode where the rectification effect in achieved via the injection locking due to a locally injected dc current $I_{AC}$ [31].

Such a behavior suggests the idea to design spintronic diodes exhibiting asymmetric rectified curves with the ascending branch as large as possible. We found that this characteristic has been already experimentally observed in [31], where a spintronic diode with a sensitivity exceeding $10^5$ V/W at zero bias field has been measured. The device structure is still an MTJ having a tilted free layer and in-plane polarizer [see inset Fig. 1(c)] which exhibits the abovementioned asymmetric rectification response (see Fig. 3(d) in Ref. [31]). Such an experimental $V_{DC} - I_{DC}$ curve reproduces a ReLU-like AF that can be described by a third-order polynomial function [see Fig. 1(c)]. For this device, we have accounted device-to-device variations by performing a linear transformation (rotation of $\theta = \pm 5°$) of the diode response, as shown in the top-left inset of Fig. 1(c). The rotation angle has been chosen by estimating the angle of the experimental data in Fig. 3(a) of [13] between the input powers 0.32 and 0.56 µW. Indeed, we assumed that the different $V_{DC} - I_{DC}$ responses to different input powers mimic the responses due to device-to-device variations.

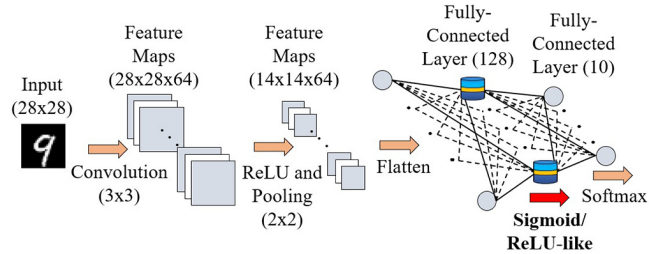

Fig. 2. Sketch of the structure of the CNN constructed for the recognition task

## III. APPLICATION IN CONVOLUTIONAL NEURAL NETWORKS

In this last part of the work, we have trained a "vanilla" convolutional neural network (**CNN**) to recognize the categories of the Mixed National Institute of Standards and Technology (**MNIST**) and Fashion-MNIST database, consisting of grayscale images 28 times size and organized in ten categories.
Fig. 2 depicts the network whose feature learning part is composed of a single convolutional layer with 64 filters of size 3×3. Its output feature maps propagate through the ReLU AF and a pooling layer that halves the results' spatial dimension with a max pooling operation. The CNN is then characterized by a flattened layer and an FC layer with 128 neurons. Finally, there is an FC layer with ten neurons (softmax AF), which returns the ten probability outputs, thus obtaining the class to which the input belongs. The first step of the experiment is the identification of the neural network parameters while considering all the neurons of the FC layer having the same AF, and to prevent the overfitting, dropout layers and early stopping are used. Second, with the weights of the ideal CNNs (all neurons of the FC layer have the same AFs), we also create 10 000 different instances of the CNN, each of them characterized by a nonideal FC layer (all the neurons of the FC have different AFs). In the following, we describe in detail the two studies performed.

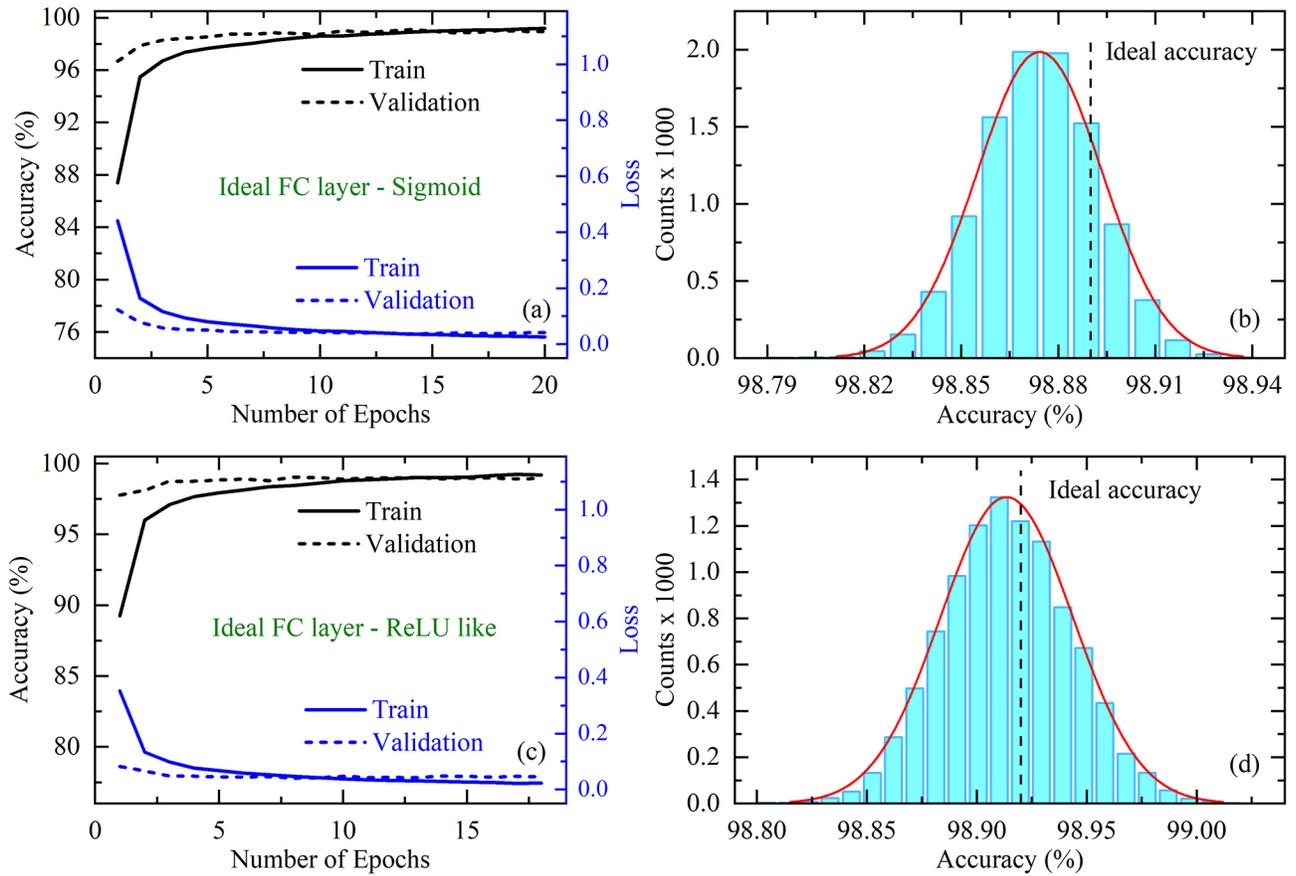

Fig. 3. (a) and (c) Accuracy versus number of epochs for training and validation achieved for neurons with ideal sigmoid and ReLU-like AFs, respectively (all neurons equals). The solid (dashed) blue line corresponds to the train (validation) loss. (b) and (d) Statistics of the accuracy obtained with the nonideal FC layer composed of spintronic neurons with different sigmoid and ReLU-like AFs. Very similar results are obtained with a fourth-order polynomial function for the fitting in Fig. 1(c).

*A. Results for the Sigmoid AF*

We train the network using the FC with ideal sigmoid AF $\alpha=1$, and we obtain, for the validation set, a percentage of accuracy of 98.93% [see Fig. 3(a)] and, for the test set, 98.89 % [ideal accuracy in Fig 3(b)]. Similarly, we obtain a percentage of accuracy in the validation set of 92.26% and in the test set of 91.81% for the Fashion-MNIST. We also found a negligible effect on the accuracy when two MTJs with different perpendicular anisotropy are used for the implementation of the AF with current.

As already explained earlier, we have used this trained network to test the effect of the device-to-device variation in the form of different magnetic anisotropy values. We set different slopes for the sigmoid, sampling 128 random values from a Gaussian distribution with a unitary mean (slope of the ideal sigmoid) and a standard deviation of 0.2 (all the negative values are discarded, and the values are resampled). We consider 10 000 realizations to build the statistics of the accuracy evaluated with the test data. Fig. 3(b) summarizes the results, as it can be seen that the accuracy can be well described by a Gaussian distribution with mean of 98.87%.

*B. Results for the ReLU-Like AF*

We also train the network using the FC with the ReLU-like AF shown in Fig. 1(c), and we obtain, for the validation set, a percentage of accuracy of 98.95% [see Fig. 3(c)] while, for the test set, 98.92% [ideal accuracy in Fig. 3(d)]. Similarly, we obtain a percentage of accuracy in the validation set of 91.93% and in the test set of 91.89% for the Fashion-MNIST. These results are comparable with the 98.87% accuracy obtained with the ideal ReLU AF, which returns the maximum between 0 and the input value. Similarly to the case of the sigmoid AF, we have also studied the effect of device-to-device variation considering AF curves between the boundary shown in the top-left inset of Fig. 1(c) ($-5° < \theta < 5°$) using a Gaussian distribution with zero mean and standard deviation of 3. The results achieved from 10 000 realizations are shown in Fig. 3(d). The values are well described by a Gaussian distribution with mean of 98.91%.

## IV. CONCLUSION

In summary, we developed device concepts for spintronic neurons based on the use of hybrid MTJs (perpendicular free layer and in-plane polarizer) where it is possible to achieve either a sigmoid or ReLU-like AF. We also studied the reliability with respect to device-to-device variations of a simple neural network, where the training is performed in an ideal scenario with all neurons having the same AF, whereas the test accuracy is computed considering the FC layer of the network with all neurons having a different AF. We obtained promising results with the test accuracy exhibiting a small variation of less than a few percentages for both sigmoid and ReLU-like AFs.

## ACKNOWLEDGMENT

This work was supported under Grant 2019-1-U.0 "Diodi spintronici rad-hard ad elevate sensitività—DIOSPIN" which is funded by the Italian Space Agency within the call "Nuove idee per la componentistica spaziale del future." The work of Riccardo Tomasello and Giovanni Finocchio was supported in part by the project ThunderSKY funded from the Hellenic Foundation for Research and Innovation and the General Secretariat for Research and Technology under Grant 871 and in part by the PETASPIN Association (www.petaspin.com).